# Commuting Network Spillovers and COVID-19 Deaths Across US Counties

Christopher Seto
Dept. of Sociology and Criminology
The Pennsylvania State University
State College, PA, United States
chs37@psu.edu

Aria Khademi
College of Information Science and Technology
The Pennsylvania State University
State College, PA, United States
khademi@psu.edu

Corina Graif
Dept. of Sociology and Criminology
The Pennsylvania State University
State College, PA, United States
corina.graif@psu.edu

Vasant G. Honavar
College of Information Science and Technology
The Pennsylvania State University
State College, PA, United States
vhonavar@ist.psu.edu

**Abstract (150 words)**

This study explored how population mobility flows form commuting networks across US counties and influence the spread of COVID-19. We utilized 3-level mixed effects negative binomial regression models to estimate the impact of network COVID-19 exposure on county confirmed cases and deaths over time. We also conducted weighting-based analyses to estimate the causal effect of network exposure. Results showed that commuting networks matter for COVID-19 deaths and cases, net of spatial proximity, socioeconomic, and demographic factors. Different local racial and ethnic concentrations are also associated with unequal outcomes. These findings suggest that commuting is an important causal mechanism in the spread of COVID-19 and highlight the significance of interconnected of communities. The results suggest that local level mitigation and prevention efforts are more effective when complemented by similar efforts in the network of connected places. Implications for research on inequality in health and flexible work arrangements are discussed.

**Introduction**

The coronavirus disease 2019 (COVID-19) pandemic has dramatically impacted societies globally, with over 32 million confirmed cases and over 980,000 COVID-19 deaths worldwide at the time of this writing[1] (Hopkins, 2020). Consequently, a growing body of research seeks to understand the social and demographic predictors of this disease at the community level, identifying local etiological factors such as age structure (Dowd et al., 2020), population density (Sy et al., 2020), and racial composition of the residents (Millett et al., 2020). In addition to local factors, equally important it is to understand the role of social contacts (Block et al., 2020) within and across communities, such as the extent to which the movement of people between communities facilitate the transmission of this infectious disease. One important type of such movement is commuting for work, a routine mobility activity that millions of people in the US engage in, typically on a daily basis (McKenzie, 2015). Many of the local and state level mitigation and prevention policies have involved some form of social distancing recommendations to "flatten the curve", in recognition that close physical proximity among people (in the regular course of their daily activities such as in the workplace, at church, or in school) can contribute significantly contributor to the spread of this disease.

Research on the transmission of this disease across space, between places such as work areas and residential areas is still in its infancy, yet important evidence is starting to emerge. For instance, (Bai et al., 2020) analyzed inter-county commuting flows in the state of New York and found that "community spreader" counties were characterized by high commuting flows to and from other counties. These findings are consistent with prior research focused on the spread of other infectious diseases which finds that commuting is an important mechanism through which diseases may be transmitted to new populations. For example, (Xu et al., 2019) linked road traffic among Chinese cities to the incidence of Influenza A (H1N1) during the 2009 pandemic. Understanding how exposures to coronavirus in an area's commuting network affects local cases and deaths is important in guiding thinking and policy in support of remote working schedule and other flexible work arrangements. Because many types of jobs do not permit remote work, certain populations, often underpaid and socioeconomically vulnerable minority groups, are disproportionately affected both at work and at home by increased risk of exposure to this disease. Moreover, these same groups are further disadvantaged disproportionately by school closures and the need to find alternative arrangements for the care of school age children and other dependents.

---

[1] As of 09.25.2020

We contribute to the extant literature on the social and spatial dynamics of COVID-19 by analyzing population across United States (US) counties which we consider to be linked via a network of commuting ties. We assess the extent to which county rates of COVID-19 deaths and cases are predicted by COVID-19 cases in linked counties, controlling for relevant structural and sociodemographic characteristics and spatial contiguity. We leverage methodological strategies from computational statistics to assess model fit and estimate significance while accounting for spatial and network dependencies within the data. Our findings demonstrate that commuting networks are an important determinant of the spread of COVID-19, as measured by deaths and confirmed cases.

**Methods**

*COVID-19 Data and Analytic Strategy*

We analyze a population of all US counties. Data on total number of COVID-19 confirmed cases and total number of COVID-19 deaths are drawn from a database maintained by USA Facts[2], which is updated daily and contains counts by county and state. We utilize 3-level mixed effects negative binomial models, analyzing COVID-19 cases and deaths of county-time periods (N=31,380), nested within counties (N=3,139), nested within states (N=51, includes DC). These models are implemented using the *menbreg* command in Stata 16 (StataCorp, 2019). Negative binomial models are well suited to predicting overdispersed count outcomes (Osgood, 2000), making them well suited to this research application. We incorporate state-level random intercepts to account for cross-state variation in COVID-19 outcomes which may have been driven by state-level policy differences (e.g., different masking requirements and enforcement of business lockdowns) and county-level random intercepts to account for unmeasured variation across counties in COVID-19 susceptibility and response. County-time periods, our first-level units of analysis, are based on the number of new COVID-19 cases and deaths for a given county within a given two weeks. Within each county are nested ten of these county-time periods, ranging from April 1st to August 18th, 2020. We use total county population (based on the 2018 American Community Survey 5-year population estimates) as an exposure term for all models, making the model coefficients interpretable as population rates. All models are estimated using Huber-White robust standard errors.

---

[2] https://usafacts.org/visualizations/coronavirus-covid-19-spread-map/

*Permutation Tests*

As a result of the network and spatial interdependencies which we hypothesize to exist among counties, conventional, analytic tests of statistical significance may fail to produce accurate confidence estimates (LeSage, 2015). Instead, we utilize a permutation testing, a flexible, simulation-based approach (Breiman, 2001; Graif et al., 2019). For each predictor, we conduct 100 permutations in which the values of the predictor are randomly permuted across all observations, breaking any association with COVID-19 mortality rates. Each permuted dataset is used to calculate model error, generating a distribution of what model error would look like if the predictor had no effect. The observed error is then compared to this distribution in order to assess the contribution which the predictor makes to model fit. A relatively low proportion of permuted cases which produced a lower error than that which was observed shows a significant contribution to model fit.

In these permutation tests, we use mean arctangent absolute percentage error (MAAPE) to capture average model error. MAAPE is computed by averaging the arctangent of the ratio of error to observed value for each observation, as shown in Equation 1. MAAPE has the advantage of capturing error as a percent, making it less sensitive to outliers than MAE, while also being robust to observations for which the true value of y is 0 (an advantage over MAPE) (Kim & Kim, 2016).

$$MAAPE = \frac{1}{N} \sum_{i=1}^{N} arctan\left(\frac{|y_i - \hat{y}_i|}{y_i}\right) \qquad (1)$$

*Network and Spatial Measures*

We used data on intercounty commuting, used to construct a weighted average of network-lagged COVID-19 exposure, were drawn from the LEHD Origin-Destination Employment Statistics (LODES) dataset, which is publicly available from the U.S. Census Bureau (US Census

– LEHD) (Graif et al., 2017; Kelling et al., 2020) . This measure was created according to Equation 2, where a given home county (h) is connected to W work counties. $C_{h-w}$ represents the number of commuters from county h who commute to county w, while $C_{h-total}$ represents the total number of outgoing commuters from county h. An additional measure of COVID-19 exposure was also created based on spatial proximity using average rate of confirmed cases of all (queen) contiguous counties (e.g., Equation 3 for a county which borders B counties). We incorporate a temporal lag into the construction of these measures by using cases from the prior two-week time period. We also incorporate measures capturing network and spatial change in COVID-19 cases from the prior to current time period which use identical weighting (Equations 4 and 5). Finally, we control for each county's own COVID-19 case rate during the prior 2 weeks.

$$\sum_{w=1}^{W} \frac{C_{h-w}}{C_{h-total}} \left( \frac{Cases_w}{Population_w} \right) (100{,}000) \qquad (2)$$

$$\sum_{b=1}^{B} \frac{1}{B} \left( \frac{Cases_b}{Population_b} \right) (100{,}000) \qquad (3)$$

$$\sum_{w=1}^{W} \frac{C_{h-w}}{C_{h-total}} \left( \frac{\Delta Cases_w}{Population_w} \right) (100{,}000) \qquad (4)$$

$$\sum_{b=1}^{B} \frac{1}{B} \left( \frac{\Delta Cases_b}{Population_b} \right) (100{,}000) \qquad (5)$$

*Causal Attribution Framework*

We used the Rubin-Neyman causal inference potential outcomes framework (Rubin, 2005) to estimate the causal effect of each of the county-level characteristics, including economic disadvantage, percentage of population over the age of 65, etc. (see Table 1 for details), on the number of deaths by COVID-19 in that county. To estimate the causal effects, we applied the well-established weighting procedure in causal inference by following a two-step mechanism: First, we weight each data sample so as to adjust for the effect of confounding and generate a weighted population that we can consider "as if randomized." Second, we perform a weighted regression where we regress total number of deaths by COVID-19 against county-level characteristics. We repeated this two-step procedure for each county-level characteristic separately, each time designating a characteristic as "treatment," and estimated the causal effect of that characteristic on total number of deaths by COVID-19.

We used the following state-of-the-art weighting methods for causal inference from observational data. Each of the three methods that we applied use a different methodology for computing the weights (in the weight model). (I) Covariate balancing propensity score weighting (CBPSW): Propensity score is defined as the probability of receiving the treatment given the covariates and is used in estimating the causal effect of binary treatments on outcomes. Propensity density is its counterpart for coping with continuous treatments. CBPSW was recently proposed and it estimates the weights based on the propensity score (for binary treatments) and propensity density (for continuous treatments) while maximizing covariate balance between the treated and controlled via an additional balancing constraint in the optimization (Fong et al., 2018; Imai & Ratkovic, 2014) (II) Inverse probability of treatment weighting: This method weights each of data samples proportionate to the inverse of the propensity score (Robins et al., 2000). (III) Super learner: This method offers a doubly robust estimate of causal effects computed through an ensemble of propensity score estimators (Pirracchio et al., 2015; Van der Laan et al., 2007). The methods that we have used have been shown to be reliable, effective, and efficient in estimating causal effects from observational data in various applications (Khademi et al., 2019; Khademi & Honavar, 2019, 2020).

The weighted outcome regression model determines the causal effect of each county level characteristic on deaths by COVID-19 through statistical hypothesis testing. We tested for the null hypothesis that each such causal effect is zero. A statistically non-significant p-value would determine a non-significant causal effect. A statistically significant p-value shows a significant causal effect and the degree (and sign) of the causal effect is determined by the magnitude (and sign) of the estimated coefficient for the treatment in the outcome regression model. We used 0.05 as the statistical significance level.

*Sociodemographic Controls*

Research has indicated that communities that have lower socioeconomic status can have more preexisting health conditions, lower access to healthcare, lower access to high-speed internet that could enable remote work, and are less able to engage in social distancing during the Covid-19 pandemic (Chiou & Tucker, 2020; Weill et al., 2020). For these reasons, several sociodemographic controls are included in the analyses. These measures were drawn from the 2014-2018 American Community Survey (ACS) 5-year estimates. Note that these measures are county-level attributes, i.e., considered to be invariant over time for the two-week time periods defining the level-one units. Economic disadvantage was measured as the first principal component produced following an analysis of unemployment rate, median income, percent in poverty, percent female-headed households, percent college graduates, percent owner-occupied housing units, and percent vacant housing units (eigenvalue = 3.2). We also include the percent of residents of 65 years or older, as well as binary indicators of whether the county is above average regarding (1) percent non-Hispanic White, (2) percent non-Hispanic Black, and (3) Hispanic. Finally, we include a measure of the percent of the county with urban residence, as measured in 2010. Table 1 shows descriptive statistics for all measures described above.

**Table 1. Descriptive statistics for entire analytic sample**

|  | Mean | SD | Min. | Max. |
|---|---|---|---|---|
| **Outcomes** | | | | |
| Total COVID-19 deaths | 5.29 | 42.57 | 0 | 2578.00 |
| Total COVID-19 confirmed cases | 166.25 | 945.21 | 0 | 41134.00 |
| **Predictors** | | | | |
| Network lagged confirmed case rate ($t_{n-1}$) | 122.43 | 129.19 | .60 | 1371.39 |
| Δ network lagged confirmed case rate ($t_{n-1} - t_n$) | 19.79 | 73.75 | -856.49 | 810.75 |
| Spatially lagged confirmed case rate ($t_{n-1}$) | 102.60 | 139.12 | 0 | 2346.52 |
| Δ spatially lagged confirmed case rate ($t_{n-1} - t_n$) | 20.06 | 98.14 | -2195.78 | 2761.80 |
| Confirmed case rate ($t_{n-1}$) | 102.26 | 209.10 | -52.25 | 13726.10 |
| Economic disadvantage | -.01 | 1.77 | -6.17 | 8.43 |
| % 65 and older | 18.37 | 4.58 | 3.80 | 55.60 |
| Above avg. NHW | .64 | --- | 0 | 1 |
| Above avg. NHB | .27 | --- | 0 | 1 |
| Above avg. Hispanic | .21 | --- | 0 | 1 |
| % Urban population | 41.34 | 31.50 | 0 | 100 |

Notes: N = 31,380 county-times nested within 3,139 counties, nested within 51 states (includes DC)

**Results**

Table 2 displays coefficient and standard error estimates from multilevel negative binomial models predicting total deaths and total confirmed cases using the full analytic sample. Results from these models are consistent with prior literature and theoretical expectations. As shown, the commuting network-based measures are robust predictors of both total deaths and total cases. This is true for both the network measure based on confirmed cases at the prior time period and the network measure capturing change in cases from the prior time period. Note that, when these network measures are taken into account, spatial contiguity is not a strong predictor of COVID-19 spread across counties. Estimated coefficients for other measures are also consistent with extant research and our expectations. Economic disadvantage, concentration of racial/ethnic minority groups, and urban population are associated with higher rates of COVID-19 cases and deaths, while a higher concentration of non-Hispanic White population is associated with lower COVID-19 cases and deaths (Tai et al., 2020). Population percent aged 65 years and older is negatively associated with cases, but positively associated with deaths (Le Couteur et al., 2020). As expected, a county's case rate at the prior time point is a strong predictor of cases at the current time point.

**Table 2. Negative binomial models (with state and county random intercepts) predicting COVID-19 outcomes across 10 time periods based on network, spatial, and time lagged cases**

|  | Total Deaths | | | Total confirmed cases | | |
|---|---|---|---|---|---|---|
|  | Beta | | SE | Beta | | SE |
| Network lagged confirmed case rate ($t_{n-1}$) | .0029 | *** | (.001) | .0037 | *** | (.001) |
| Δ network lagged confirmed case rate ($t_{n-1} - t_n$) | .0007 | * | (.000) | .0044 | *** | (.000) |
| Spatially lagged confirmed case rate ($t_{n-1}$) | .0003 | | (.000) | .0003 | | (.000) |
| Δ spatially lagged confirmed case rate ($t_{n-1} - t_n$) | .0003 | † | (.000) | .0004 | † | (.000) |
| Confirmed case rate ($t_{n-1}$) | .0018 | *** | (.000) | .0010 | *** | (.000) |
| Economic disadvantage | .0471 | ** | (.015) | .0285 | ** | (.009) |
| % 65 and older | .0184 | ** | (.007) | -.0285 | *** | (.005) |
| Above avg. NHW | -.3510 | *** | (.060) | -.2164 | *** | (.041) |
| Above avg. NHB | .1963 | *** | (.058) | .1933 | ** | (.062) |
| Above avg. Hispanic | .0680 | | (.068) | .3074 | *** | (.055) |
| % Urban population | .0061 | *** | (.001) | .0025 | *** | (.001) |
| Two-week time period (ref. 4/1-4/14) | | | | | | |
|     4/15-4/28 | .0268 | | (.063) | .0295 | | (.038) |
|     4/29-5/12 | .0718 | | (.063) | .0424 | | (.045) |
|     5/13-5/26 | -.1195 | † | (.065) | .1300 | * | (.052) |
|     5/27-6/9 | -.1840 | ** | (.070) | .1958 | *** | (.056) |
|     6/10-6/23 | -.4006 | *** | (.077) | .3480 | *** | (.079) |
|     6/24-7/7 | -.5674 | *** | (.138) | .4828 | *** | (.077) |
|     7/8-7/21 | -.7162 | *** | (.123) | .6501 | *** | (.085) |
|     7/22-8/4 | -.5829 | *** | (.140) | .7632 | *** | (.100) |
|     8/5-8/18 | -.4671 | *** | (.136) | .8684 | *** | (.103) |
| Constant | -11.9601 | *** | (.181) | -7.9921 | *** | (.107) |
| ln(alpha) | -.4932 | *** | (.073) | -.6243 | *** | (.051) |
| State-level variance | .3184 | *** | (.088) | .0971 | *** | (.030) |
| County-level variance | .4817 | *** | (.047) | .1993 | *** | (.019) |

Notes: Exposure = County race-specific population 2014-2018, ACS 5-year estimates; NHB = Non-Hispanic Black; ***p < .001; ** p < .01; * p < .05; † p < 0.10

The permutation test results shown in Table 3 provide further support for these findings. The p-values shown in Table 3 are based on the proportion of trials in which a model based on a random permutation of each respective variable outperformed the original model based on the observed data, as measured by MAAPE. Based on this alternative measure of significance, the network-based measures still emerge as very salient predictors, i.e., these measures meaningfully improve model fit. Specifically, the network-based measures improved model fit in all trials for the models predicting deaths and cases (i.e., no model in which this measure was permuted outperformed the original model based on observed data).

Table 3. Negative binomial models (with state and county random intercepts) predicting COVID-19 outcomes across 10 time periods based on network, spatial, and time lagged cases: Permutation test results

|  | Total deaths | | Total confirmed cases | |
| --- | --- | --- | --- | --- |
|  | Beta | Proportion | Beta | Proportion |
| Network lagged confirmed case rate ($t_{n-1}$) | .0029 | .00 | .0037 | .00 |
| Δ network lagged confirmed case rate ($t_{n-1}$ - $t_n$) | .0007 | .00 | .0044 | .00 |
| Spatially lagged confirmed case rate ($t_{n-1}$) | .0003 | .43 | .0003 | .01 |
| Δ spatially lagged confirmed case rate ($t_{n-1}$ - $t_n$) | .0003 | .10 | .0004 | .00 |
| Confirmed case rate ($t_{n-1}$) | .0018 | .00 | .0010 | .00 |
| Economic disadvantage | .0471 | .00 | .0285 | .24 |
| % 65 and older | .0184 | .03 | -.0285 | 1.00 |
| Above avg. NHW | -.3510 | 1.00 | -.2164 | .95 |
| Above avg. NHB | .1963 | .99 | .1933 | 1.00 |
| Above avg. Hispanic | .0680 | .20 | .3074 | .40 |
| % Urban population | .0061 | 1.00 | .0025 | 1.00 |

Notes: Exposure = County race-specific population 2014-2018, ACS 5-year estimates; NHB = Non-Hispanic Black; ***p < .001; ** p < .01; * p < .05; † p < 0.10

*Sensitivity Tests*

We conducted several tests to assess how findings changed with alternative model specifications. In order to better separate network effects from possible unmeasured spatial confounders, we re-estimated the models using network measures based on (1) only contiguous counties and (2) only non-contiguous counties. Tables 4 and 5 show estimates from these models (respectively). As shown, the network effects persist in both cases, further supporting our finding that commuting networks matter for the spread of COVID-19 beyond spatial proximity.

**Table 4. Negative binomial models (with state and county random intercepts) predicting COVID-19 outcomes across 10 time periods based on network, spatial, and time lagged cases. Network based on only contiguous counties**

|  | Total Deaths | | | Total confirmed cases | | |
|---|---|---|---|---|---|---|
|  | Beta | | SE | Beta | | SE |
| Network lagged confirmed case rate ($t_{n-1}$) | .0016 | *** | (.000) | .0027 | *** | (.000) |
| Δ network lagged confirmed case rate ($t_{n-1}$ - $t_n$) | .0006 | *** | (.000) | .0028 | *** | (.000) |
| Spatially lagged confirmed case rate ($t_{n-1}$) | .0009 | * | (.000) | .0004 | | (.000) |
| Δ spatially lagged confirmed case rate ($t_{n-1}$ - $t_n$) | .0004 | * | (.000) | .0007 | * | (.000) |
| Confirmed case rate ($t_{n-1}$) | .0019 | *** | (.000) | .0011 | *** | (.000) |
| Economic disadvantage | .0407 | ** | (.015) | .0262 | ** | (.009) |
| % 65 and older | .0196 | ** | (.007) | -.0270 | *** | (.005) |
| Above avg. NHW | -.3323 | *** | (.059) | -.2020 | *** | (.042) |
| Above avg. NHB | .1920 | *** | (.056) | .1878 | *** | (.057) |
| Above avg. Hispanic | .0670 | | (.068) | .3110 | *** | (.055) |
| % Urban population | .0063 | *** | (.001) | .0028 | *** | (.001) |
| Two-week time period (ref. 4/1-4/14) | | | | | | |
|     4/15-4/28 | .0649 | | (.058) | .0075 | | (.036) |
|     4/29-5/12 | .1035 | † | (.061) | .0323 | | (.051) |
|     5/13-5/26 | -.0903 | | (.068) | .0850 | | (.059) |
|     5/27-6/9 | -.1643 | * | (.072) | .1546 | * | (.068) |
|     6/10-6/23 | -.3869 | *** | (.083) | .3365 | *** | (.100) |
|     6/24-7/7 | -.5375 | *** | (.133) | .6002 | *** | (.110) |
|     7/8-7/21 | -.6019 | *** | (.127) | .8233 | *** | (.112) |
|     7/22-8/4 | -.4160 | * | (.167) | .8870 | *** | (.123) |
|     8/5-8/18 | -.3349 | * | (.157) | .9290 | *** | (.124) |
| Constant | -11.9509 | *** | (.186) | -7.9040 | *** | (.109) |
| ln(alpha) | -.4609 | *** | (.077) | -.5884 | *** | (.056) |
| State-level variance | .3206 | *** | (.082) | .1006 | *** | (.027) |
| County-level variance | .4696 | *** | (.046) | .1917 | *** | (.018) |

Notes: Exposure = County race-specific population 2014-2018, ACS 5-year estimates; NHB = Non-Hispanic Black; ***p < .001; ** p < .01; * p < .05; † p < 0.10

To aid our causal inference, we also conducted several analyses using different weighting strategies on a cross-sectional version of our data in which outcomes are cumulative counts of a county's cases or deaths, and network and spatially lagged measures are based on these cumulative counts. Results from these models are shown in Table 6. As shown, these alternative model specifications produced substantively similar results with regard to the commuting network effects, offering further support for our conclusions. Results of all of the causal effect estimators consistently show that the percentage of population over the age of 65 and economic

**Table 5. Negative binomial models (with state and county random intercepts) predicting COVID-19 outcomes across 10 time periods based on network, spatial, and time lagged cases. Network based on only non-contiguous counties.**

|  | Total Deaths | | | Total confirmed cases | | |
|---|---:|---|---|---:|---|---|
|  | Beta | | SE | Beta | | SE |
| Network lagged confirmed case rate ($t_{n-1}$) | .0020 | *** | (.001) | .0020 | *** | (.001) |
| Δ network lagged confirmed case rate ($t_{n-1} - t_n$) | .0004 | | (.000) | .0029 | *** | (.000) |
| Spatially lagged confirmed case rate ($t_{n-1}$) | .0013 | *** | (.000) | .0019 | *** | (.000) |
| Δ spatially lagged confirmed case rate ($t_{n-1} - t_n$) | .0006 | *** | (.000) | .0022 | *** | (.000) |
| Confirmed case rate ($t_{n-1}$) | .0019 | *** | (.000) | .0011 | *** | (.000) |
| Economic disadvantage | .0407 | ** | (.015) | .0219 | * | (.009) |
| % 65 and older | .0178 | ** | (.007) | -.0296 | *** | (.005) |
| Above avg. NHW | -.3566 | *** | (.059) | -.2207 | *** | (.040) |
| Above avg. NHB | .1820 | ** | (.058) | .1716 | ** | (.062) |
| Above avg. Hispanic | .0779 | | (.068) | .3276 | *** | (.055) |
| % Urban population | .0062 | *** | (.001) | .0026 | *** | (.001) |
| Two week time period (ref. 4/1-4/14) | | | | | | |
|     4/15-4/28 | .0136 | | (.067) | .0514 | | (.040) |
|     4/29-5/12 | .0564 | | (.068) | .0714 | | (.046) |
|     5/13-5/26 | -.1305 | † | (.067) | .1552 | ** | (.052) |
|     5/27-6/9 | -.1949 | ** | (.072) | .2130 | *** | (.057) |
|     6/10-6/23 | -.4107 | *** | (.078) | .3662 | *** | (.080) |
|     6/24-7/7 | -.5675 | *** | (.141) | .5164 | *** | (.078) |
|     7/8-7/21 | -.7066 | *** | (.127) | .6784 | *** | (.087) |
|     7/22-8/4 | -.5932 | *** | (.142) | .7961 | *** | (.110) |
|     8/5-8/18 | -.5011 | *** | (.135) | .8969 | *** | (.116) |
| Constant | -11.9407 | *** | (.180) | -7.9712 | *** | (.115) |
| ln(alpha) | -.4806 | *** | (.072) | -.5905 | *** | (.051) |
| State-level variance | .3090 | *** | (.087) | .1007 | *** | (.031) |
| County-level variance | .4873 | *** | (.048) | .2042 | *** | (.021) |

Notes: Exposure = County race-specific population 2014-2018, ACS 5-year estimates; NHB = Non-Hispanic Black; ***p < .001; ** p < .01; * p < .05; † p < 0.10

disadvantage have significant and considerable causal effects on the total number of deaths by COVID-19. The results of these causal analyses demonstrate the vulnerability of elderly and those economically disadvantages to COVID-19.

**Table 6. Estimates and statistical significance of the causal effects of county-level characteristics on deaths by COVID-19.**

*Covariate Balancing Propensity Score Weighting (CBPSW)*

| | Estimate | SE | Z-value | P-value |
|---|---|---|---|---|
| Network lagged confirmed case rate | .003 | .000 | 45.937 | <0.001 |
| Economic disadvantage | .282 | .112 | 2.512 | .012 |
| Percent 65 and older | .438 | .009 | 46.881 | <0.001 |
| At least 25% NHB | -2.095 | .084 | -25.040 | .000 |
| At least 25% Hispanic | .003 | .000 | 18.067 | .000 |
| Percent urban population | .031 | .001 | 36.574 | .000 |
| Spatially lagged confirmed case rate | .001 | .000 | 4.129 | .000 |

*Inverse Probability of Treatment Weighting (IPTW)*

| | Estimate | SE | Z-value | P-value |
|---|---|---|---|---|
| Network lagged confirmed case rate | .003 | .000 | 13.391 | .000 |
| Economic disadvantage | .353 | .054 | 6.590 | .000 |
| Percent 65 and older | .438 | .009 | 47.331 | <0.001 |
| At least 25% NHB | -2.100 | .084 | -25.015 | .000 |
| At least 25% Hispanic | .003 | .000 | 18.285 | .000 |
| Percent urban population | .031 | .001 | 38.371 | <0.001 |
| Spatially lagged confirmed case rate | .001 | .000 | 8.497 | .000 |

*Super Learner*

| | Estimate | SE | Z-value | P-value |
|---|---|---|---|---|
| Network lagged confirmed case rate | .003 | .000 | 13.132 | .000 |
| Economic disadvantage | .335 | .076 | 4.418 | .000 |
| Percent 65 and older | .440 | .009 | 46.767 | <0.001 |
| At least 25% NHB | -2.100 | .090 | -23.403 | .000 |
| At least 25% Hispanic | .003 | .000 | 17.861 | .000 |
| Percent urban population | .031 | .001 | 42.346 | <0.001 |
| Spatially lagged confirmed case rate | .001 | .000 | 6.739 | .000 |

**Discussion**

This study found that an area's population exposure to COVID-19 in the area's commuting network contributes to higher local levels of confirmed COVID –19 cases and deaths, above and beyond the area's socioeconomic disadvantage, age composition, urban status, and racial and ethnic composition. Importantly, the local effect of network exposure to COVID –19 cases is robust to controlling for spatial contiguity to COVID-19 cases exposure, and to controlling for prior local levels of COVID-19 exposures. The causal effect estimates of the network and spatial lag variables are consistent with the previous estimates. These results have important implications for future research and policy. They showed that, during this pandemic, places have been significantly affected by their residents' exposures to infection through their commuting networks across the country. This indicates that policies that focus on local level mitigation and prevention efforts are more effective when complemented by similar efforts in the many extra-local connected places across the states and the country.

Given the growing research suggesting that vulnerable populations are less able to work remotely and engage in physical distancing during this pandemic, our results also indicate the acute need for work level protections, such as providing paid sick days, increasing minimal wages, providing health safety equipment to essential workers, to assisting with childcare for working parents who have to work while the schools are closed or in remote mode. These necessary provisions will not only help save the lives and health of workers who cannot afford to socially distance themselves from their work environments, but they have the great potential to spillover and improve the fates of whole communities that their workers go back home to.

As expected, an area's socioeconomic disadvantage contributed to both higher death rates and cases relative to the local population. The area's concentration of whites was associated with a protective effect against both infection cases and COVID-19 deaths. The concentration of minorities, both above average share of Hispanics and non-Hispanic Blacks was associated with higher rates of confirmed cases, consistent with a large body of work that has documented the many challenges associated with COVID-19 risk that burden minority communities, including the higher likelihood to be in frontline occupations and in other low paid occupations that have little flexibility and cannot be easily be transitioned to remote work format.[3]

*Limitations.*

These data come with several limitations. The analytical focus was on counties in part due to restrictions regarding the COVID-19 data availability across the country. To the extent that the data access and granularity expands in future months, analyses at more local levels will be very valuable. Still, analyses on other important transmittable diseases like influenza have examined place-to-place transmission patterns for geographic units as large as states and counties (Bozick & Real, 2015) with important lessons that have inspired further research.

The network measures used in this study were limited by the data access constraints to information updated on an annual basis, and thus they do not capture the fast-occurring changes during this ongoing pandemic. While these measures captured the commuting network prior the pandemic, those links have likely been weakened by layoffs or remote work transitions. Still, the information on the COVID-19 rates within the commuting network was captured as it changed over time. Given that the pandemic likely contributed to weakening rather than strengthening pre-existing commuting links across places, the fact that nevertheless we still see strong effects suggests to us that adjustments in the future to these data to reflect the rapid changes in employment status will likely reveal even stronger effects of commuting exposures to COVID-19.

---

[3] https://www.brookings.edu/research/to-protect-frontline-workers-during-and-after-covid-19-we-must-define-who-they-are/

*Implications and Future Directions.*

Many businesses across the country have restricted their employment during the COVID-19 pandemic, some have even closed temporarily or permanently, while others allowed employees to work remotely for the purpose of "social distancing" and in the hope of "flattening the curve" (Bartik et al., 2020). Understanding how these mobility changes and restrictions contribute to containing the COVID-19 transmission is an important next step for future research. Moreover, it is known that some population groups are more likely to be in occupations (e.g., health care providers, grocery workers, bus drivers, meatpacking workers) that have been on the frontlines in the fight against COVID-19, unable to comply with social distancing recommendations and policies. Understanding how workplace networks and risk transmission differentially affect disadvantaged and minority populations is of great importance in future research. Importantly, also understanding the types of workplace connections and other social network-based distancing strategies (Block et al., 2020) that can work best to contain the pandemic risk without further isolating the most vulnerable populations and communities is essential.